%% file: main.tex
\documentclass[12pt]{article}
\usepackage[letterpaper,top=2cm,bottom=2cm,left=3cm,right=3cm]{geometry}
\usepackage{graphicx}
\usepackage{amsmath,amssymb,amsfonts}
\usepackage{amsthm}
\usepackage{booktabs,multirow}
\usepackage[dvipsnames]{xcolor}
\usepackage[colorlinks=true,allcolors=blue]{hyperref}
\usepackage{cleveref}
\usepackage{array}
\usepackage{authblk}
\usepackage[
        range-phrase=--,
        range-units=single,
        retain-explicit-plus=true,
        retain-unity-mantissa=false,
    ]{siunitx}

\newcommand{\maxsqz}{\SI{-0.81}{dB}}
\newcommand{\maxasqz}{\SI{+4.29}{dB}}
\newcommand{\maxsqzerr}{\SI{0.04}{dB}}
\newcommand{\maxasqzerr}{\SI{0.10}{dB}}

\newcommand{\pump}{\SI{27}{mW}}
\newcommand{\maxsqzonchip}{\SI{-7.52}{dB}}
\newcommand{\maxasqzonchip}{\SI{+9.62}{dB}}
\newcommand{\maxsqzonchiperr}{\SI{0.22}{dB}}
\newcommand{\maxasqzonchiperr}{\SI{0.25}{dB}}

\newcommand{\visible}{\SI{793.5}{nm}}
\newcommand{\nir}{\SI{1587}{nm}}
\newcommand{\sqzbandwidth}{\SI{10.3}{THz}}

\theoremstyle{plain}

\theoremstyle{definition}

\theoremstyle{remark}

\newcommand*\samethanks[1][\value{footnote}]{\footnotemark[#1]}

\title{Quantum squeezing in an all-resonant periodically poled lithium niobate microresonator}

\author[1,2]{Xinyi Ren\thanks{These authors contributed equally.}}
\author[1,2]{Reshma Kopparapu\samethanks}
\author[1,4]{Tushar Sanjay Karnik\samethanks}
\author[1,2]{Chun-Ho Lee\samethanks}
\author[1]{Kiwon Kwon}
\author[1,2]{Clayton Cheung}
\author[1,2,4]{Yue Yu}
\author[3]{Shi-Yuan Ma}
\author[3]{Bo-Han Wu}
\author[1,2]{Ran Yin}
\author[1,2]{Lian Zhou}
\author[2]{Quntao Zhuang}
\author[3]{Dirk Englund}
\author[1,2]{Zaijun Chen}
\author[1,2,4]{Mengjie Yu\thanks{Corresponding author: \href{mailto:mengjie.yu@berkeley.edu}{mengjie.yu@berkeley.edu}}}

\affil[1]{Department of Electrical Engineering and Computer Sciences, University of California, Berkeley, CA 94720, USA}
\affil[2]{Ming Hsieh Department of Electrical and Computer Engineering, University of Southern California, Los Angeles, CA 90089, USA}
\affil[3]{Research Laboratory of Electronics, MIT, Cambridge, MA 02139, USA}
\affil[4]{Materials Sciences Division, Lawrence Berkeley National Laboratory, Berkeley, CA 94720, USA}

\date{} % no date on arXiv

\begin{document}
% \linenumbers
\maketitle

\begin{abstract}
Quantum noise limits the sensitivity of optical measurements, but squeezed states of light enable quantum-enhanced metrology, sensing, and information processing. Most on-chip squeezed-light sources rely on Kerr ($\chi^{(3)}$) nonlinearities, remain limited by pump power and excess loss constraints. Quadratic ($\chi^{(2)}$) platforms instead provide stronger parametric interactions, lower pump power requirements, and greater spectral engineering flexibility. Here, we demonstrate strong, broadband squeezed-light generation on a thin-film lithium niobate (TFLN) photonic chip using a dual-resonant optical parametric amplifier implemented in a single periodically poled LN (PPLN) microresonator. Near-full-depth domain inversion is achieved simultaneously with highly over-coupled resonances, exhibiting escape efficiencies exceeding 90$\%$  and intrinsic quality factors above 2.5 million in a $\SI{0.6}{mm}^2$ X-cut TF-PPLN resonator, enabling efficient squeezing at \nir{} when pumped at \visible{}. Operating in the continuous-wave regime, we directly measure \maxsqz{} of squeezing below the shot-noise limit with a pump power of \pump{}, together with \maxasqz{} of anti-squeezing. From these measurements, we infer an on-chip squeezing level of $\maxsqzonchip{} \pm \maxsqzonchiperr{}$ (95\% confidence interval: $[-7.96,-7.10]$~dB), and an on-chip anti-squeezing level of $\maxasqzonchip{} \pm \maxasqzonchiperr{}$ . 
We demonstrate single-mode squeezing at degeneracy with a squeezed-light spectrum exceeding \sqzbandwidth{}. This work reports the highest squeezing ratio among integrated $\chi^{(2)}$ cavity platforms and the first quasi-phase-matched, fully resonant $\chi^{(2)}$ cavity squeezer on chip, establishing a scalable route to fully integrated power-efficient squeezed-light sources for quantum-enhanced sensing and metrology.
\end{abstract}

\section{Introduction}\label{sec1}
Squeezed states of light~\cite{yuen1976two,walls1983squeezed}, in which quantum noise in one field quadrature is suppressed below the shot-noise limit, are indispensable resources for continuous-variable quantum technologies. By reducing measurement uncertainty beyond the standard quantum limit, squeezing has enabled breakthroughs in quantum sensing~\cite{lawrie2019quantum}, including acoustic frequency sensing~\cite{novikov_hybrid_2025}, gravitational-wave detection~\cite{jia_squeezing_2024, ganapathy_broadband_2023, virgo_collaboration_frequency-dependent_2023, zhao_frequency-dependent_2020}, quantum imaging~\cite{casacio_quantum-enhanced_2021, taylor_biological_2013}, dual-comb spectroscopy~\cite{herman_squeezed_2025,hariri2025} and distributed quantum sensing~\cite{zhang2021distributed,xia2020demonstration,guo2020distributed,xia2023entanglement}, as well as applications in quantum communication and secure key distribution~\cite{fesquet_demonstration_2024, nguyen_digital_2025}. Furthermore, squeezed vacuum and multimode squeezed combs~\cite{guidry_multimode_2023, wang_large-scale_2025} underpin many architectures for continuous-variable quantum computation and simulation~\cite{konno_logical_2024, aghaee_rad_scaling_2025}. Although bulk optical parametric oscillators (OPOs) have long provided squeezing~\cite{vahlbruch_detection_2016, schnabel_squeezed_2017}, these systems are large-scale, alignment-sensitive, and power-intensive, limiting their practicality in scalable quantum networks. A major goal of modern photonics is therefore to realize compact and scalable squeezed light sources directly on photonic chips. Integrated photonic platforms offer compelling advantages: they can leverage strong optical nonlinearities in tightly confined waveguides and microresonators to dramatically reduce the required pump power and to achieve broadband squeezing bandwidth via powerful dispersion engineering~\cite{hoff_integrated_2015,dutt_-chip_2015}.

Over the past decade, motivated by these advantages, integrated photonics has emerged as a versatile platform for squeezed light generation, encompassing both quadratic $\chi^{(2)}$ and Kerr $\chi^{(3)}$ nonlinearities, as well as waveguide- and resonator-based implementations operating in pulsed and continuous-wave (CW) regimes. 
Early demonstrations of on-chip squeezing were achieved in silicon nitride microring resonators based on Kerr nonlinearity~\cite{dutt_-chip_2015}, followed by substantial development in oscillation threshold reduction, bandwidth extension, and near-degenerate operation~\cite{lu_milliwatt-threshold_2019, zhao_near-degenerate_2020, vaidya_broadband_2020,zhang_squeezed_2021, yang_squeezed_2021,riemensberger_photonic_2022, shen_strong_2025,shen_highly_2025,jahanbozorgi_generation_2023,Cernansky_SPM_2020,wang_large-scale_2025}.
Despite these advances, Kerr-based platforms face intrinsic limitations. Owing to the higher-order nature of the Kerr interaction, achieving strong squeezing typically requires high pump power or extreme cavity quality ($Q$) factors, which in turn introduce broadband excess noise or parasitic nonlinear processes, such as thermo-refractive fluctuations and spontaneous Raman scattering~\cite{Cernansky_SPM_2020, vaidya_broadband_2020}. In resonant devices, the close spectral proximity of the pump and squeezed fields enforces unfavorable trade-offs between escape efficiency and pump power, as overcoupling the squeezing resonance rapidly increases the required pump power~\cite{yang_squeezed_2021, jahanbozorgi_generation_2023}. This property further necessitates high–extinction-ratio pump filtering for homodyne detection, introducing additional optical loss that directly degrades observable squeezing~\cite{yang_squeezed_2021, jahanbozorgi_generation_2023, zhao_near-degenerate_2020, zhang_squeezed_2021, dutt_-chip_2015}. In addition, competing nonlinear processes such as non-degenerate four-wave mixing and Bragg scattering can be readily phase-matched and, when driven at high pump power, lead to parasitic mode coupling and excess noise that degrade squeezing purity and operational robustness~\cite{zhao_near-degenerate_2020, vaidya_broadband_2020, zhang_squeezed_2021}.

These challenges have motivated increasing interest in quadratic ($\chi^{(2)}$) nonlinear platforms, where the substantially stronger interaction enables efficient parametric processes at reduced pump power. In practice, integrated $\chi^{(2)}$ squeezed-light generation has been realized almost exclusively in lithium niobate, owing to its large second-order nonlinearity, low optical loss, and quasi-phase-matching technology. Thin-film lithium niobate (TFLN) further combines these material advantages with high optical confinement and scalable nanofabrication, making it a leading platform for integrated continuous-variable quantum photonics. Ultra-broadband pulsed vacuum squeezing has also been demonstrated in nanophotonic $\chi^{(2)}$ waveguides, highlighting the strong nonlinear capability of this platform~\cite{nehra_few-cycle_2022}.
Initial demonstrations of squeezed-light generation in TFLN have primarily relied on non-resonant periodically poled lithium niobate (PPLN) waveguides, which require large device footprints and are pumped by either pulsed or continuous-wave lasers at high optical power. While such platforms enable terahertz-bandwidth squeezing, they often require tens to hundreds of milliwatts of pump power~\cite{mondain_chip-based_2019, chen_ultra-broadband_2022, shi_squeezed_2025, Kashiwazaki_wg_2020} and are susceptible to waveguide damage and laser noise. Moreover, achieving uniform quasi-phase matching over centimeter-scale interaction lengths remains technically challenging, as poling nonuniformity directly degrades squeezing efficiency and spectral purity. Cavity-enhanced TFLN approaches offer a promising route to lowering the pump threshold by resonantly enhancing the nonlinear interaction while reducing the device footprint; however, simultaneously achieving strong light–matter interaction through high-fidelity poling and large Purcell enhancement, while optimizing the resonance conditions and coupling strengths for both the pump and squeezed modes, remains a central challenge for realizing robust, low-threshold continuous-wave squeezing on chip~\cite{bruch_-chip_2019, lu_periodically_2019, lu_ultralow-threshold_2021}. To date, squeezing achieved in TFLN resonators include an integrated OPO demonstrating \SI{0.55}{dB} squeezing and \SI{1.55}{dB} anti-squeezing, where only the fundamental field is resonant with \SI{35}{\%} escape efficiency~\cite{park_single-mode_2024} or using modal phase matching with \SI{1.5}{dB} inferred squeezing~\cite{arge_demonstration_2025}. 

In this work, we demonstrate quadrature squeezing, for the first time, using an all-resonant quasi-phase-matched optical parametric amplifier fabricated on the TFLN. Leveraging advanced nanofabrication and device co-design, we realize a TF-PPLN microresonator with the highest intrinsic quality factor of 2.6 million reported to date for a PPLN cavity, together with near-full-depth domain inversion and a poling duty cycle approaching 50\%, all achieved in an extremely over-coupled resonance implemented with a 200-µm coupling length. This enables strong and reproducible resonant enhancement of the $\chi^{(2)}$ interaction.
Operating below threshold at a pump power of \pump{} and precise thermal control of a co-resonance condition, we observe a noise suppression of $\maxsqz{} \pm \maxsqzerr{}$ relative to shot noise, together with $\maxasqz{} \pm \maxasqzerr{}$ of anti-squeezing. We infer an on-chip squeezing of $\maxsqzonchip{} \pm \maxsqzonchiperr{}$ and anti-squeezing of $\maxasqz{} \pm \maxasqzerr{}$ from both loss tracking and photon flux methods. The generated squeezed light exhibits a broadband spectrum with a squeezing bandwidth of \sqzbandwidth{}, corresponding to 244 correlated signal–idler mode pairs and an on-chip squeezed-light power of \SI{320}{pW} at degeneracy.

\section{Results}\label{sec2}
%%%%%%%%%%%%%%%%%%%%%%%%
\begin{figure*}[t]
    \centering
    \includegraphics[width=\textwidth]
    {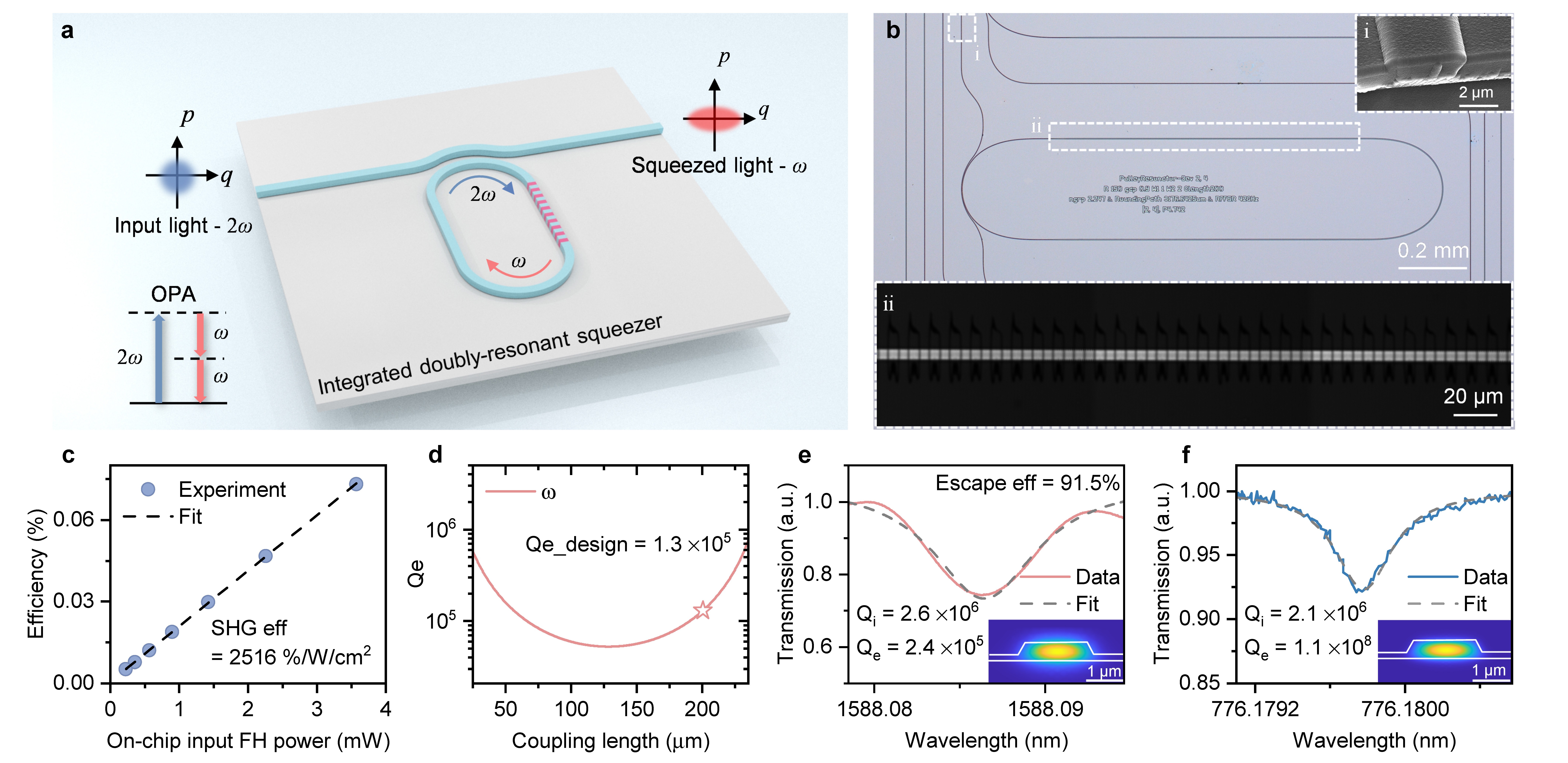}
    \caption{Concept and design of the integrated dual-resonant on-chip squeezer based on TF-PPLN. 
    \textbf{a}, Schematic of the microring resonator with PPLN on a nanophotonic chip. The device generates squeezed light at the FH frequency $\omega$ (red, output) from a SH frequency at $2\omega$ (blue, input) through a $\chi^{(2)}$ nonlinear interaction. 
    \textbf{b}, Optical micrograph of the microring with the \SI{0.9}{mm}-long periodical-poled section highlighted. A SEM image of the anti-reflection–coated chip facet is shown on the top right. The bottom panel shows its microscope image of the poled domain pattern inside the resonator. 
    \textbf{c}, SHG efficiency calibration of a nearby poled waveguide, yielding \SI{2516}{\%/W/cm^2}. 
    \textbf{d}, Pulley coupler design illustrating phase matching of TE$_0$ modes between bus and resonator waveguides. The FH mode is intentionally over-coupled to enable efficient external coupling. 
    \textbf{e,f}, Experimental transmission spectra at FH (\textbf{e}) and SH (\textbf{f}) wavelengths, showing agreement with the coupled-mode design with the extracted external $Q$-factors of $Q_\mathrm{e} = 2.4 \times 10^{5}$, while intrinsic $Q$-factors are $Q_\mathrm{i} = 2.1 \times 10^{6}$ (SH) and $Q_\mathrm{i} = 2.6 \times 10^{6}$ (FH). The over-coupled FH resonance yields an escape efficiency of $\sim 91.5\%$. Insets show the simulated spatial mode profiles, confirming that both FH and SH modes are supported within the same waveguide, enabling dual-resonant operation.
    }
    \label{fig:concept}
\end{figure*}
%%%%%%%%%%%%%%%%%%%%%%%%

The squeezed vacuum is generated on a photonic integrated circuit (PIC) featuring a single-ring resonator equipped with PPLN, as shown in~\cref{fig:concept}a. The microring is designed to be resonant at both the near-793.5-nm second-harmonic (SH) and near-1587-nm fundamental-harmonic (FH) wavelengths, thereby enabling efficient $\chi^{(2)}$ optical parametric interaction. 
The racetrack resonator is fabricated on a 600-nm-thick X-cut TFLN wafer with an etch depth of \SI{350}{nm} after periodic poling on one straight waveguide arm and followed by cladding with a 800-nm-thick silicon dioxide layer. The cavity path length of \SI{3177}{\micro\metre} corresponds to a free spectral range (FSR) of approximately \SI{42.2}{GHz} (\cref{fig:concept}a). Due to the device dispersion and finite cavity linewidth,  only a subset of cavity resonances pairs across the FH and SH spectral bands satisfy co-resonance condition where both light with frequency $\omega$ and $2\omega$ can simultaneously be resonantly enhanced. In~\cref{fig:concept}b, the top panel shows a microscope image of the microring resonator. The region marked by white box~$i$ corresponds to the scanning electron microscope (SEM) image of chip facets, which are coated with optimized anti-reflection layers to suppress Fresnel reflections and parasitic etalon effects for the squeezing light output. 

Quasi-phase-matching (QPM) is achieved through electric-field-assisted periodic poling with a period of \SI{4.742}{\micro\metre}, chosen to generate the squeezed light at wavelength within the \SIrange{1540}{1610}{nm} band based on the device dimension. White box~$ii$ highlights the region designated for periodic poling. High-voltage pulses with a few milliseconds in duration, are applied across lithographically defined electrodes to invert ferroelectric domains prior to waveguide etching. The bottom panel of~\cref{fig:concept}b shows a portion of the \SI{0.9}{mm}-long poled section. By processing the 2-photon microscope image, we estimate a poling duty cycle of 0.54 $\pm$ 0.01 and an intensity contrast of 1.03 $\pm$ 0.04 between the poled and unpoled sections (see supplementary material SM Sec. II) . Based upon these inferred values, our simulations predict a second-harmonic generation (SHG) efficiency of \SI{2570}{\%/W/cm^2} which is in good agreement with the experimentally measured value of  \SI{2516}{\%/W/cm^2} obtained from an adjacent poled waveguide (\cref{fig:concept}c).

The optical pulley coupler in the racetrack resonator is designed using dielectric perturbation theory. Along the ring waveguide path, the waveguide width is varied: the straight sections at the top and bottom are set to \SI{2}{\micro\metre} to enhance nonlinear mode confinement and reduce loss, while tapering adiabatically to a narrower width of \SI{1}{\micro\metre} when entering the curved sections with Euler bends. This design enables efficient phase matching between the TE$_0$ modes of the bus and ring waveguides while suppressing coupling to undesired higher-order modes, and provides over-coupling at the FH wavelength to maximize the escape efficiency of the squeezed light. As shown in \cref{fig:concept}d, the pulley-coupler is designed to yield an external $Q$-factor $Q_\mathrm{e} \approx 1.3 \times 10^{5}$ for the FH mode (star marker). The measured resonances closely match this design. From the FH transmission spectrum in \cref{fig:concept}e, we extract an external $Q$-factor of $Q_\mathrm{e} = 2.4 \times 10^{5}$, indicating a strongly overcoupled cavity and yielding an escape efficiency of $\eta_{\mathrm{esc}} = Q_{\mathrm{i}}/(Q_{\mathrm{i}} + Q_{\mathrm{e}}) = 91.5\%$. The intrinsic $Q$-factors extracted from \cref{fig:concept}(e,f) are $Q_\mathrm{i} = 2.6 \times 10^{6}$ for the FH mode and $Q_\mathrm{i} = 2.1 \times 10^{6}$ for the SH mode. 
To our knowledge, it is the highest escape efficiency and $Q_{\mathrm{i}}$ ever achieved in the TF-PPLN resonators, as a result of precise coupling engineering and co-design with low-loss nanofabrication processes (see SM Sec. I). In previously reported on-chip resonators, pushing toward strong overcoupling often leads to substantial degradation of intrinsic quality factor due to parasitic coupling to undesired modes and additional scattering loss introduced in the coupling region~\cite{pfeiffer_coupling_2017, puckett_422million_2021}. In a resonant squeezing platform, high escape efficiency ensures that the majority of the generated nonclassical field is extracted into the measurement channel rather than dissipated internally, making it a key requirement for achieving large observable squeezing.

%%%%%%%%%%%%%%%%%%%%%%%%
\begin{figure*}[t]   % the * makes it span both columns
    \centering
    \includegraphics[width=\textwidth]
    {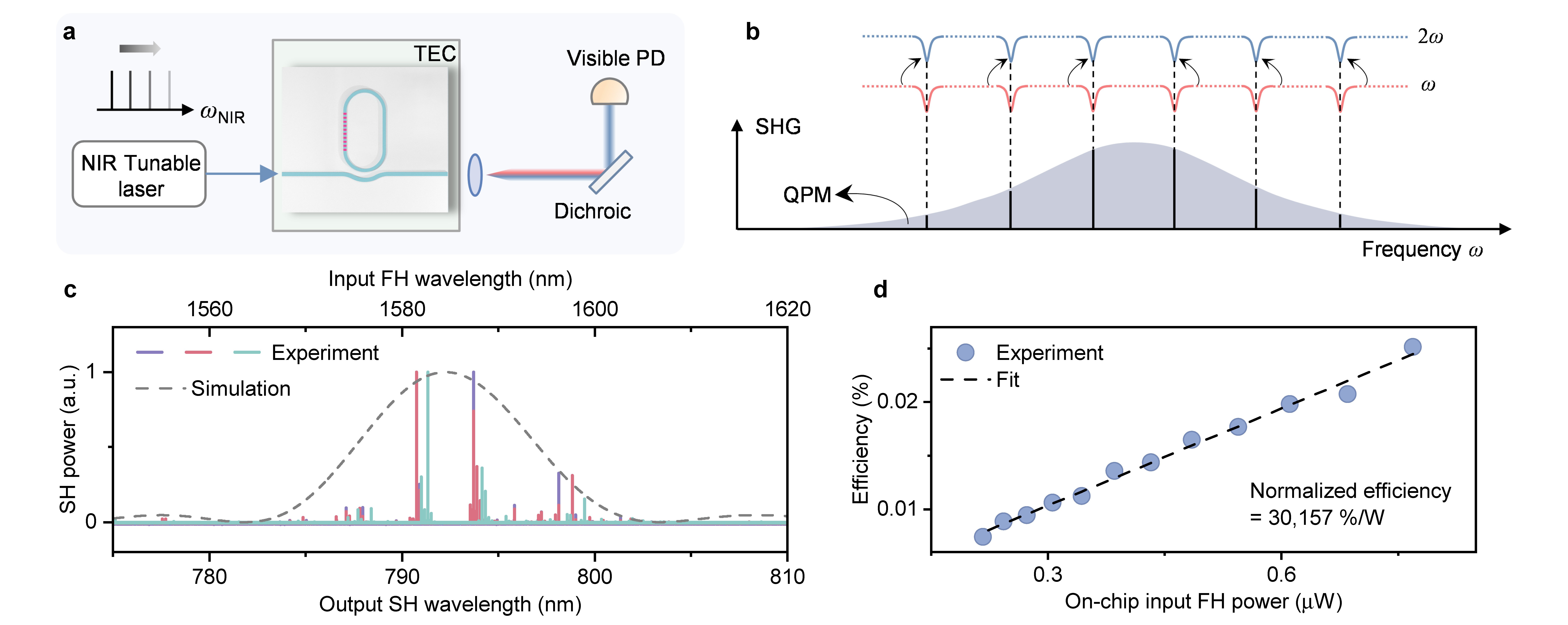}
    \caption{SHG characterization of the on-chip $\chi^{(2)}$ squeezer.
    \textbf{a}, SHG measurement setup. A tunable NIR laser is coupled into the squeezer chip with the generated visible light collected by a visible PD. 
    \textbf{b}, SHG peaks appear at discrete positions within the QPM bandwidth, where the FH and SH resonances simultaneously align. The peak spacing of approximately \SI{4}{\nano\metre} arises from the resonator FSRs of the two modes. 
    \textbf{c}, SHG spectra measured at different temperatures with the overall envelope matching with the simulated QPM response of the poled waveguide.
    \textbf{d}, Power-dependent SHG efficiency showing a normalized on-chip conversion efficiency of \SI{30157e4}{\%/\mathrm{W}}, extracted from a linear fit to experimental data.
    }
    \label{fig:SHGcharacterization}
\end{figure*}
%%%%%%%%%%%%%%%%%%%%%%%%

We first characterize the nonlinear performance of our squeezer chip by measuring its SHG spectrum using the setup shown in \cref{fig:SHGcharacterization}a (see SM Sec. III for details).  The chip is mounted on a thermoelectric controller (TEC) for temperature tuning and stabilization. Although the SHG occurs across the entire QPM bandwidth of \SI{20}{nm} (\SI{3}{dB}) centered around \SI{1587}{nm}, only the light of the wavelength satisfying the co-resonance condition can build up inside the cavity and consequently produce a significant SH output. As shown in~\cref{fig:SHGcharacterization}b, several discrete SHG peaks appear within the QPM range, with a spacing determined by the resonator FSRs of the FH and SH modes. This co-resonance condition can be further optimized by temperature, exploiting the different thermo-optic coefficients of the FH and SH modes, as shown in Figure~\ref{fig:SHGcharacterization}c.
Despite these changes in the discrete spectral peaks, the global spectral envelope remains the same and closely matches the simulated QPM response of the poled waveguide, confirming that the underlying nonlinear interaction and domain structure are uniform. The strong agreement between the measured envelope and the model further verifies the high fidelity of the periodic poling. The simulated envelope is obtained from a first-order quasi-phase-matched waveguide model using the numerically computed effective-index dispersion $n_{\mathrm{eff}}(\lambda)$ of the guided mode. We calculate the phase mismatch $\Delta k(\lambda)=k_{2\omega}(\lambda/2)-2k_{\omega}(\lambda)$, include the grating wavevector $2\pi/\Lambda$ associated with the uniform poling period $\Lambda$, and evaluate the normalized spectral response $\propto \mathrm{sinc}^2\!\left([\Delta k(\lambda)-2\pi/\Lambda]L/2\right)$ for a uniform interaction length $L=\SI{0.9}{\milli\meter}$. This model assumes undepleted pump, single-mode propagation, and uniform domain structure, and therefore captures the global quasi-phase-matching envelope rather than the discrete resonant mode structure.
Figure~\ref{fig:SHGcharacterization}d shows the SHG efficiency calibration performed under the co-resonant condition. The visible pump wavelength is selected at the peak of the measured SH spectrum. In the undepleted-pump regime, the generated second-harmonic power increases linearly with the on-chip fundamental power. A linear fit to the data yields a normalized SHG efficiency of \SI[parse-numbers=false]{30,157}{\%/W}.

\begin{figure*}[t]   % the * makes it span both columns
    \centering
    \includegraphics[width=\textwidth]
    {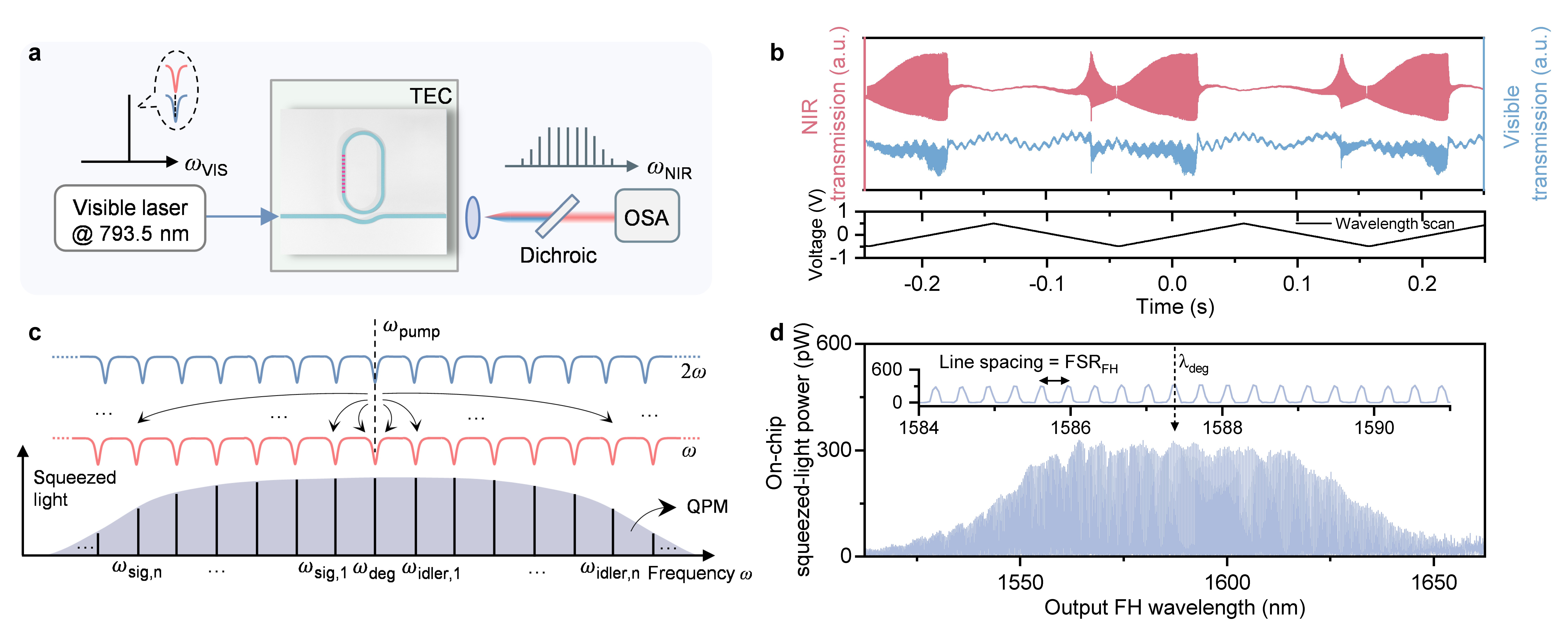}
    \caption{
    Squeezed light spectrum characterization of the on-chip $\chi^{(2)}$ squeezer. 
    \textbf{a}, Measurement setup. A visible pump laser near \visible drives the squeezer chip under the dual-resonant condition. An OSA records the output spectrum. 
    \textbf{b}, Transmission spectra of the two modes recorded, showing alternating amplification and de-amplification features when the resonances overlap.
    \textbf{c}, When the pump excites the co-resonant SH and FH modes, both degenerate and nondegenerate mode pairs are generated, producing a broadband squeezed light spectrum with evenly spaced sidebands. 
    \textbf{d}, Measured on-chip quantum frequency comb spectrum, revealing broadband parametric generation around \nir. The inset shows zoomed-in cavity resonances with uniform spacing matching the FSR$_{\mathrm{FH}}$.
    }
    \label{fig:OPAcharacterization}
\end{figure*}

To characterize the parametric emission process, we first inject both SH and FH light and monitor their transmission while tuning the chip temperature to locate the co-resonance condition (\cref{fig:OPAcharacterization}a and b, also see SM Sec. IV). \cref{fig:OPAcharacterization}b shows the NIR (FH) and visible (SH) transmission spectra under optimal temperature tuning, corresponding to near-zero relative detuning between the two modes. The NIR transmission fringes are a result of a seeded phase-sensitive parametric amplification and de-amplification. The FH input is then blocked and only the SH pump—near \visible—is launched into the cavity. The resulting squeezed light spectrum is recorded using an optical spectrum analyzer (OSA). 
The underlying parametric process and the experimental results are illustrated in~\cref{fig:OPAcharacterization}c and~\cref{fig:OPAcharacterization}d, respectively. Both degenerate and nondegenerate parametric sidebands are generated, producing a broadband squeezed-light spectrum with evenly spaced frequency components. The inset in~\cref{fig:OPAcharacterization}d shows the frequency spacing matches the NIR FSR (FSR$_{\mathrm{FH}}$) of \SI{42.2}{GHz}. The spontaneous degenerate parametric sidebands are achieved at \nir{} with an on-chip power of \SI{320}{pW} over a cavity linewidth of \SI{1.36}{GHz}, corresponding to 2.05 photon number per Hz which can be also used to infer on-chip squeezing (see SM Sec. VII). Furthermore, the output spontaneous parametric spectrum features more than \SI{85}{nW} from \SI{1512}{nm} to \SI{1655}{nm} and a 3-dB bandwidth over \sqzbandwidth{}  consisting of 244 mode pairs. The high brightness and the large spectral coverage of the parametric spectrum are a result of the high escape efficiency, as well as the low group velocity dispersion (GVD, $\beta_2$) and the compact poling length enabled by our dispersion-engineered PPLN resonator, respectively. The platform allows for both near-degenerate single-mode and non-degenerate two-mode squeezing operation in the same device simply by configuring only the local oscillators.
Leveraging the large Purcell enhancement factor in a cavity, the resonant squeezer breaks the trade-off between squeezing ratio and bandwidth typically present in a single-pass waveguide based device, though the squeezing comes at a form of discrete frequency pairs. We further analyze the squeezed light spectra under different pump–cavity detuning conditions and find that the squeezing power and the parametric gain are maximized at a near-zero detuning (SM Sec. VI). Such a dual-resonant operation is therefore critical for achieving a strong parametric interaction and underpins the squeezing measurements discussed in subsequent sections.

%%%%%%%%%%%%%%%%%%%%%%%%
%%%%%%%%%%%%%%%%%%%%%%%%

%%%%%%%%%%%%%%%%%%%%%%%%
\begin{figure}[t]  
\centering
    \includegraphics[width=0.6\textwidth]
    {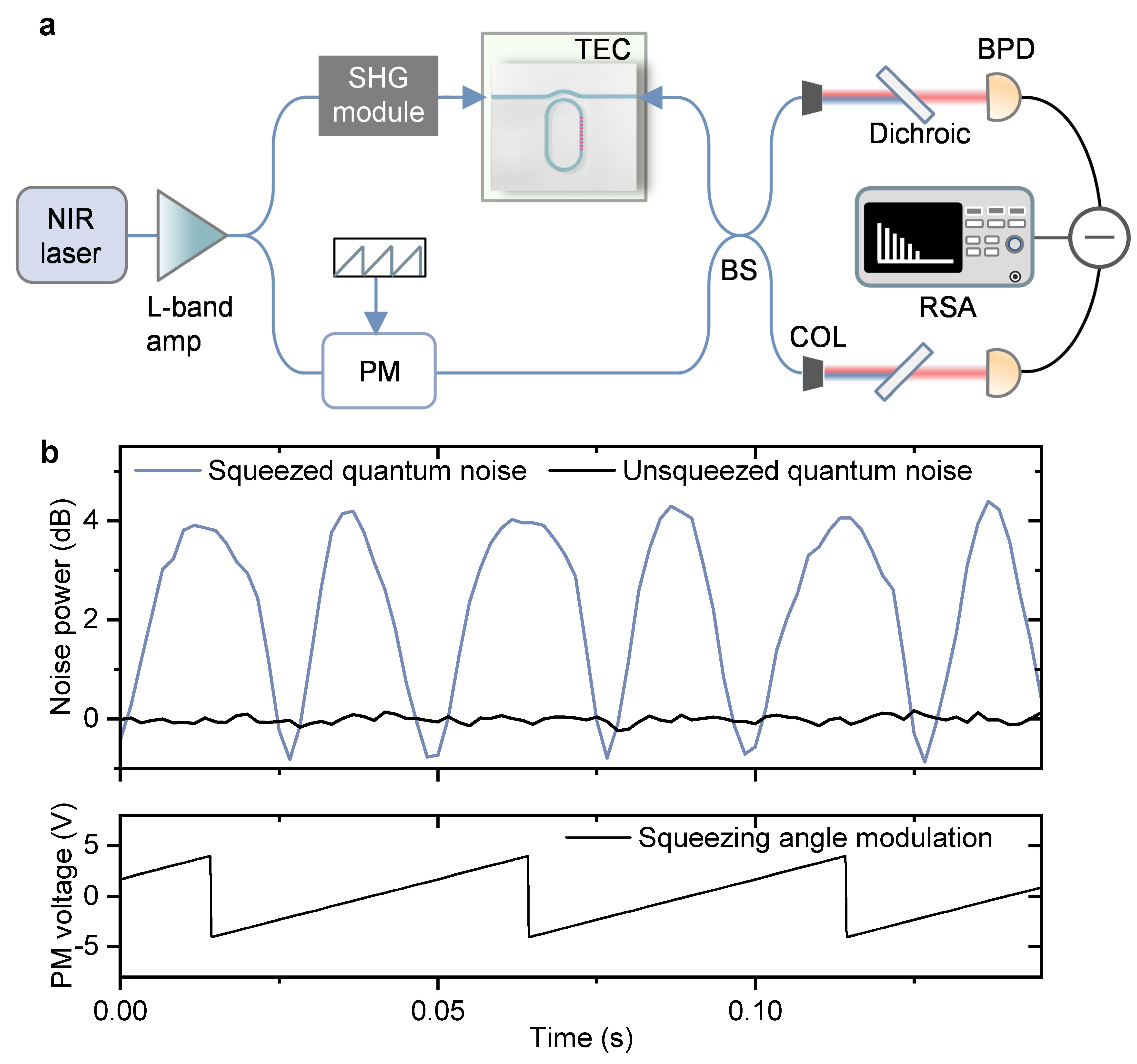}
    \caption{Observation of squeezing from the dual-resonant integrated squeezer. 
    \textbf{a}, Layout of the experimental setup. The NIR laser is split into two paths of squeezing generation and local oscillator before merging at the balanced homodyne detector. NIR: near-infrared; LO: local oscillator; PM: phase modulator; COL: collimator; BS: beam splitter; RSA: radio-frequency spectrum analyzer. 
    \textbf{b}, Normalized quantum noise under the sweeping of the squeezing angle, as shown in the lower trace of the phase-modulator drive. $\maxsqz{} \pm \maxsqzerr{}$ squeezing and $\maxasqz{} \pm \maxasqzerr{}$ anti-squeezing are observed.}
    \label{fig:squeezing}
\end{figure}
%%%%%%%%%%%%%%%%%%%%%%%%

To observe quadrature squeezing and measure the corresponding reduction in quantum noise, the chip output is then interfered with a strong local oscilator (LO) using a balanced photodetector with a measured external quantum efficiency of \SI{96}{\%} (\cref{fig:squeezing}a). The fiber-based interferometer yields a balanced-homodyne interference with \SI{17}{milliradian} relative phase stability and $\SI{91.5}{\%}$ visibility. Phase control between the squeezed vacuum and the LO is implemented using an electro-optic phase modulator (Thorlabs, LN65S-FC) in the LO path. Sweeping the squeezing angle over 0-$\pi$ enables the readout of the squeezed and anti-squeezed quadratures, corresponding to a reduced and enhanced quantum noise. In our experiment, the triangular scan from $\SI{-4}{V}$ to $\SI{+4}{V}$ corresponds to a full $2\pi$ phase sweep at \SI{20}{Hz}. \cref{fig:squeezing}b shows the detected noise power at a sideband frequency of \SI{20}{MHz}, normalized such that the shot noise corresponds to \SI{0}{dB}. We observe $\maxsqz{} \pm \maxsqzerr{}$ of squeezing and $\maxasqz{} \pm \maxasqzerr{}$ of anti-squeezing at the input on-chip pump power of \pump{}. The dark noise of the photodetector is \SI{20}{dB} lower than the shot noise level, and the total detection loss from the chip output to the BPD is \SI{6.84}{dB}. 
While the observed squeezing is primarily limited by the off-chip loss, the on-chip squeezing could also be further improved by increasing the visible pump power to approach the limit of \SI{-10.6}{dB} posted by the device escape efficiency if operating near the OPO threshold. We infer the on-chip squeezing level using two complementary analyses based on loss tracking and photon-flux estimation (see SM Sec. VII). In the loss-tracking analysis (SM Sec. VIIA), the measured squeezing and anti-squeezing are related to the on-chip values through the output facet transmission and the independently measured off-chip losses, including propagation loss, mode-overlap visibility, and photodiode quantum efficiency. Solving the resulting coupled nonlinear equations yields an output facet loss of \SI{4.44}{dB}, and on-chip squeezing and anti-squeezing levels of \maxsqzonchip{} and \maxasqzonchip{}, respectively. The extracted output facet loss implies slightly asymmetric coupling losses between the input and output facets, since the independently measured total input–output transmission loss is \SI{9}{dB}. Notably, this result demonstrates that quadrature variance measurements provide a practical method to extract the output facet loss of photonic integrated circuits, which is otherwise difficult to access directly. Additionally, we estimate the on-chip squeezing using a photon-flux model based on the measured squeezed-light power at degeneracy (SM Sec. VIIB), obtaining \SI{-7.56}{dB} and \SI{+9.70}{dB} for squeezing and anti-squeezing, respectively, in good agreement with the loss-tracking analysis. Uncertainty analysis is performed via Monte Carlo error propagation based on the measurement uncertainties of the relevant physical parameters, including the measured squeezing ($\pm \SI{0.04}{dB}$), anti-squeezing ($\pm \SI{0.10}{dB}$), effective off-chip transmission ($\pm \SI{0.2}{dB}$), and escape efficiency ($\pm \SI{0.04}{dB}$). This yields inferred on-chip squeezing and anti-squeezing values of $\maxsqzonchip \pm \maxsqzonchiperr{}$ with 95\% confidence interval of $[-7.96,-7.10]$~dB and $\maxasqzonchip \pm \maxasqzonchiperr{}$ with 95\% confidence interval of $[9.13,10.12]$~dB, respectively (details see SM Sec. VIIA)

\begin{figure}[h!]
\begin{center}
    \includegraphics[width=0.5\textwidth]{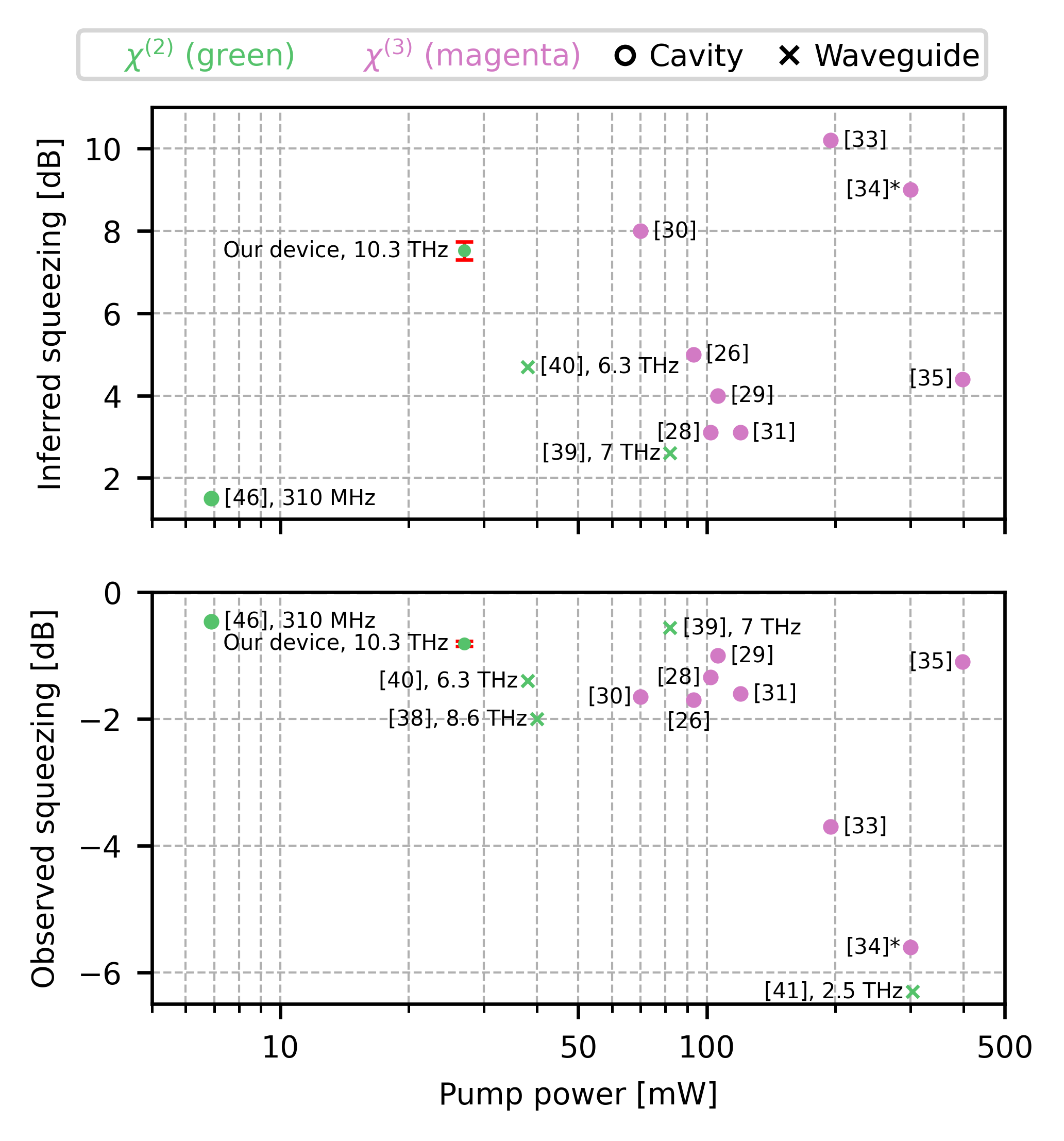}
    \caption{Literature comparison of integrated squeezed-light sources. Reported squeezing and anti-squeezing levels are plotted against on-chip pump power for integrated devices based on $\chi^{(2)}$ (green) and $\chi^{(3)}$ (magenta) nonlinearities. Marker shapes denote the device architecture of cavity (circle) and waveguide (cross). Preprint result is indicated by an asterisk. Where available, reported squeezing or squeezed light spectrum bandwidths are annotated next to the corresponding data points, including multi-terahertz bandwidths for waveguide-based devices and a \sqzbandwidth{} spectral span for our device. Our work occupies the low-power, high-squeezing regime near the left side of the plot, underscoring the advantage of dual-resonant TFLN devices for efficient squeezed-light generation. Ref.~[45] %with an on-chip pump power of \SI{4}{mW} 
    is not included due to no inferred value (measured \SI{-0.55}{dB} squeezing and \SI{1.55}{dB} anti-squeezing).
    }
    \label{fig:literature}
\end{center}
\end{figure}

Compared to the state-of-art integrated squeezed light generators, our dual-resonant thin-film lithium niobate squeezer operates at pump power well below all previously demonstrated $\chi^{(3)}$ integrated squeezers, achieves the highest squeezing ratio and broadest parametric spectrum reported for cavity-based $\chi^{(2)}$ devices, and simultaneously offers a far smaller footprint than waveguide-based implementations (\cref{fig:literature}). High-fidelity quasi-phase matching enables access to the largest effective $\chi^{(2)}$ nonlinearity in lithium niobate and supports a broader squeezing bandwidth compared with modal phase-matching approaches. When combined with simultaneous resonance at both the pump and squeezed wavelengths, this architecture dramatically reduces the required pump power, thereby mitigating the impact of excess laser noise and the risk of optical-induced damage. In contrast to $\chi^{(3)}$-based demonstrations, which require ultrahigh optical quality factors to reach appreciable squeezing, our platform achieves large squeezing at a relaxed cavity $Q$ of only a few million, substantially easing fabrication tolerances and improving scalability. The use of distinct pump and squeezing wavelengths further allows independent control of the resonant conditions, enabling operation in an extremely overcoupled regime exclusively at the squeezing wavelength while maintaining optimal pump enhancement. Moreover, this work constitutes the first demonstration to simultaneously overcome the high optical loss typically associated with domain-inversion-induced scattering, parasitic mode coupling, and etching loss in long coupling regions, while preserving strong nonlinear interaction. The resulting compact, dispersion-engineered cavity supports an extended squeezing bandwidth, providing a scalable route toward two-mode squeezing protocols spanning bandwidths exceeding \SI{10}{THz}.

%%%%%%%%%%%%%%%%%%%%%%%%
%%%%%%%%%%%%%%%%%%%%%%%%

\section{Conclusion}\label{sec3}
In conclusion, we have demonstrated a chip-scale, dual-resonant $\chi^{(2)}$ thin-film PPLN squeezer that achieves continuous-wave squeezing and anti-squeezing of $\maxsqz{} \pm \maxsqzerr{}$ and $\maxasqz{} \pm \maxasqzerr{}$ ($\maxsqzonchip{} \pm \maxsqzonchiperr{}$ and $\maxasqzonchip{} \pm \maxasqzonchiperr{}$ inferred on chip) at unprecedentedly low pump power of \pump{} while maintaining high escape efficiency of $91.5\%$, compact footprint of 0.6 mm$^{2}$, and engineered spectral support of over \SI{10}{THz} within a single integrated device. 
%The large spectral separation between the pump and squeezed fields enables low-loss pump rejection and direct homodyne characterization of the squeezing spectrum, allowing unambiguous verification of single-mode squeezing at degeneracy and fundamentally relaxing pump-filtering requirements compared with Kerr-based integrated squeezers.
Looking ahead, dispersion engineering enabled by deeper etching and air cladding will reduce the GVD and thus further extend the bandwidth of the squeezing spectrum, supporting ultra-broadband multi-terahertz two-mode squeezing protocols. At the same time, poling-related etching loss can be further reduced using etch-and-sidewall-poled fabrication approaches that avoids anisotropic etching induced scattering~\cite{Franken_UV_SPLN_2025}. The platform is intrinsically compatible with electro-optic phase control, enabling fast modulation, Pound–Drever–Hall locking, and long-term stabilization, and can be readily integrated with on-chip interferometric circuits. At the same time, continued reduction of propagation and coupling losses through optimized etching, sidewall smoothing, and low-loss fiber–chip interfaces will directly increase the detectable squeezing. Reaching operation near the parametric oscillation threshold—required for maximum gain and squeezing—will benefit from operation at elevated temperature to suppress photorefractive–thermo-optic competition, providing a pathway toward near-quantum-limited performance. Beyond squeezed-light generation, the dual-resonant architecture naturally supports low-noise parametric amplification when operated in an overcoupled regime~\cite{Zhao_regenerativeOPA_2023}, quantum frequency combs~\cite{shi_entanglementcomb_2023, kues_onchip_2017} as well as high-brightness, high-rate spontaneous parametric down-conversion sources using the same device geometry~\cite{Ma_UltrabrightQuantumPhotonSources_2020}. Together, these capabilities position thin-film lithium niobate as a scalable and multifunctional platform for compact, power-efficient, and broadband continuous-variable quantum photonics.

\section*{Data availability}
The datasets generated and analyzed in the current study are available
from the corresponding authors on reasonable request.

\section*{Acknowledgements}
This work is supported by the DARPA INSPIRED program (HR001123S0052) and the DARPA Young Faculty Award (D23AP00252-02). We thank Wenxuan Jia for technical advice on the fiber interferometer and phase noise analysis, Ian Christen for advice on experimental stabilization, and Kamila Kunes for squeezing simulations.
Two-photon imaging experiments were conducted at the CRL Molecular Imaging Center, RRID: SCR017852, supported by NIH grant S10OD025063.
Device fabrication was performed at the John O’Brien Nanofabrication Laboratory at University of Southern California and  Marvell Nanofabrication Laboratory at University of California, Berkeley.
M.Y., Y.Y. and T.S.K. are supported by the U.S. Department of Energy, Office of Science, Basic Energy Sciences, Materials Sciences and Engineering Division under Contract No. DE-AC02-05CH11231 within the Quantum Coherent Systems Program KCAS26.
The views, opinions and/or findings expressed are those of the authors and should not be interpreted as representing the official views or policies of the Department of Defense or the U.S. Government.

\section*{Author contributions}
M.Y. conceived the idea. 
C.-H.L. designed the chip with the help of X.R., R.K. and T.K.. 
C.-H.L. and T.S.K. fabricated the devices and developed the fabrication processes with the help of C.C. and Y.Y..
X.R. and R.K. carried out the experiments and analyzed the data with help from K.K., R.Y., %I.C.
and L.Z..
S.-Y.M. and B.-H.W. performed theoretical calculations, supervised by D.E..
Q.Z. developed the theoretical model for inferred squeezing.
X.R., R.K. and T.K. wrote the manuscript with contribution from all authors. 
M.Y. and Z.C. supervised the project.

\section*{Competing interests}
C.-H.L., L.Z., Z.C. and M.Y. are involved in developing lithium niobate technologies at Opticore Inc.

%%===========================================================================================%%
% \bibliographystyle{unsrt}
% \bibliography{main}% common bib file
\input{output.bbl}

\end{document}

%% file: output.bbl
\providecommand{\noopsort}[1]{}\providecommand{\singleletter}[1]{#1}%